\documentstyle[epsf,twocolumn,prl,aps,floats]{revtex}
\newcommand{\figwidthsmall}{3.0in} 
\begin{document}
\draft

 \twocolumn[\hsize\textwidth\columnwidth\hsize\csname @twocolumnfalse\endcsname

\title{ Theory of the Reentrant Charge-Order Transition in the Manganites }
\author{ C. Stephen Hellberg$^\dag$}
\address{ Center for Computational Materials Science,
Naval Research Laboratory, Washington, DC 20375}
\date{\today}
\maketitle
\begin{abstract}
\noindent
A theoretical model for the reentrant charge-order
transition in the manganites is examined.
This transition is studied with a purely 
electronic model for the $e_g$ Mn electrons: the extended 
Hubbard model. The electron-phonon coupling results in a 
large nearest-neighbor repulsion between $e_g$ electrons.
Using a finite-temperature Lanczos technique,
the model is diagonalized on a 16-site periodic cluster to calculate the 
temperature-dependent phase boundary between the charge-ordered 
and homogeneous phases.
A reentrant transition is found.
The results are discussed with respect to the specific topology of the 16-site cluster.
\end{abstract}
\pacs{
}

 ]
\makeatletter
\global\@specialpagefalse
\def\@oddhead{Submitted to the 8th Joint MMM-Intermag conference\hfill C. Stephen Hellberg}
\let\@evenhead\@oddhead

The manganites have a very rich phase diagram
that includes ferromagnetic, antiferromagnetic,
and charge-ordered phases \cite{schiffer95a,ramirez97a,mori98a}.
Various theoretical models have been used to explain different aspects
of this phase diagram \cite{moreo99a,jackeli00a,khomskii00a,brink99a}.

In its simplest incarnation, the 
charge-ordered (CO) phase occurs at hole doping $x=1/2$
with equal amounts of 
Mn$^{3+}$
and
Mn$^{4+}$
ordered in real space in a checker-board pattern.
The oxygens relax away from the
Mn$^{3+}$ ions
and towards the
Mn$^{4+}$ ions, thus providing
a repulsive potential between
Mn$^{3+}$ ions (or equivalently between
Mn$^{4+}$ ions).
The potential energy gain exceeds the kinetic energy loss
due to the formation of this insulating state \cite{vojta99a}.

When observed, the CO is generally the lowest temperature phase,
but recently the CO phase has been seen to melt with
decreasing temperature
in Pr$_{0.65}$(Ca$_{0.7}$Sr$_{0.3}$)$_{0.35}$MnO$_3$ \cite{tomioka97a} and
LaSr$_2$Mn$_2$O$_7$ \cite{kimura98a}.
The lowest temperature phase is metallic,
and the CO insulator is only observed at intermediate temperatures.
A reentrant transition has been obtained theoretically using
extended Hubbard models both with electron-phonon interactions \cite{yuan99a}
and without electron-phonon interactions \cite{pietig99a}.

In this paper, we study the charge-order
transition in the extended Hubbard model (without electron-phonon interactions)
on the two-dimensional square lattice.
Previous work \cite{pietig99a} solved this model
in infinite spatial dimensions,
resulting in finite entropy (due to the spins) at $T=0$ in the CO phase, so
a reentrant transition was guaranteed to be found.
In the infinite two-dimensional square lattice,
the spins will order into a N\'eel state with zero entropy at $T=0$.

The Hamiltonian is given by
\begin{equation}
 H = t \sum_{\langle i j \rangle \sigma } (c_{ i \sigma}^\dagger c_{ j \sigma} 
+ {\rm h.c.})
+ U \sum_i n_{i\uparrow} n_{i\downarrow}
+ V \sum_{\langle i j \rangle } n_i n_j,
\label{ham}
\end{equation}
where $c_{ i \sigma}^\dagger$ ($c_{ i \sigma}$) creates (annihilates) 
an electron with spin $\sigma$ on site $i$,
$n_{i \sigma}$ is the number operator with spin $\sigma$ on site $i$,
and $n_i = n_{i\uparrow} + n_{i\downarrow}$.
The hopping amplitude is $t$,
$\langle i j \rangle$ enumerates nearest neighbor sites
on the two-dimensional square lattice,
$U$ is onsite repulsion, and
$V$ is the nearest-neighbor repulsion.
The non-interacting bandwidth on the two-dimensional square lattice
is given by $W=8|t|$;
we set $U=W$ and vary $V$ at quarter filling (one electron 
for every two sites).
For small V, we expect the ground state will be a homogeneous
Fermi liquid.
For large V, the electrons will crystallize in a checkerboard pattern to avoid
occupying neighboring sites.

We solve the Hamiltonian (\ref{ham}) on a 4$\times$4 cluster
using a recently developed finite-temperature Lanczos technique 
\cite{hellberg99cavo,hellberg00lanczos}.
We choose periodic boundary conditions resulting in a closed Fermionic shell
in the non-interacting limit \cite{hellberg00gfmc}.
In each symmetry sector, we perform N$_{\rm L}=4000$ Lanczos steps.
We eliminate spurious eigenvalues \cite{cullum85a},
leaving more than 2000 real eigenvalues in each sector,
but not all of these will be converged.
The extreme eigenvalues (lowest and highest) converge first 
\cite{hellberg00lanczos}.
From these eigenvalues, we compute the susceptibility
\begin{equation}
\chi = \frac{1}{N} \frac{\partial E}{\partial V},
\label{susc}
\end{equation}
where $N=16$ is the number of sites.
The susceptibility $\chi$ is equivalent to the nearest-neighbor
pair correlation function
by the Kubo formula.
In the homogeneous phase $\chi$ is large, but in the CO phase
where there is only a small probability of finding electrons
on neighboring sites, $\chi$ is small.

\begin{figure}[tbh]
\epsfxsize=\figwidthsmall\centerline{\epsffile{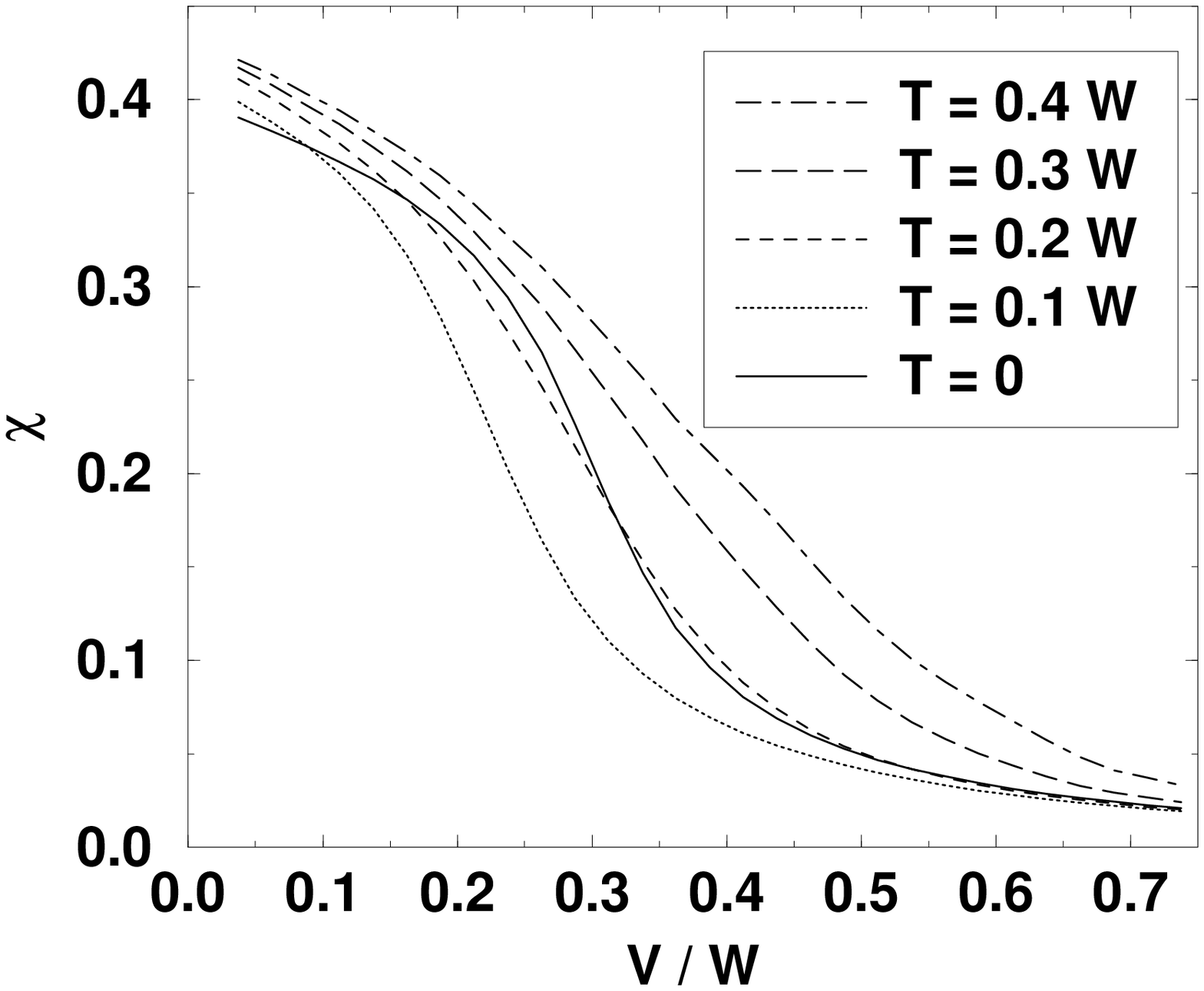}}
\vspace{.1in}
\caption{Temperature dependence of the susceptibility as a
function of the nearest-neighbor repulsion $V$ (in units
of the bandwidth $W=8|t|$) for the 4$\times$4 periodic cluster at $U=W$.
The susceptibility varies continuously even at $T=0$ (solid curve).
The susceptibility is large in the homogeneous phase (small $V$)
and small in the CO phase (large $V$).
With increasing temperature, the susceptibility first shifts to
smaller $V$ then to larger $V$, indicating a reentrant transition.
}
\label{fig:susc}
\end{figure}

The susceptibility (\ref{susc}) for five different temperatures
is plotted in Fig.\ \ref{fig:susc}.
For all temperatures shown, $\chi$ is large at small $V$ in the homogeneous
phase, decreases continuously with increasing $V$, and flattens out
at a small value in the CO phase at large $V$.
To determine the boundary between the homogeneous and CO phases,
we pick a critical value of $\chi_c = 0.1$ for the phase boundary.
We choose this value to coincide approximately with the maximum
of $\partial^2 \chi / \partial V^2$.
A different value of $\chi_c$ results in a
phase boundary that is qualitatively the same: for all
reasonable values of $\chi_c$, the phase boundary first shifts
to smaller $V$ as the temperature is raised from zero and then
moves to larger $V$ in the high-temperature regime, characteristic of a
reentrant transition.

\begin{figure}[tbh]
\epsfxsize=\figwidthsmall\centerline{\epsffile{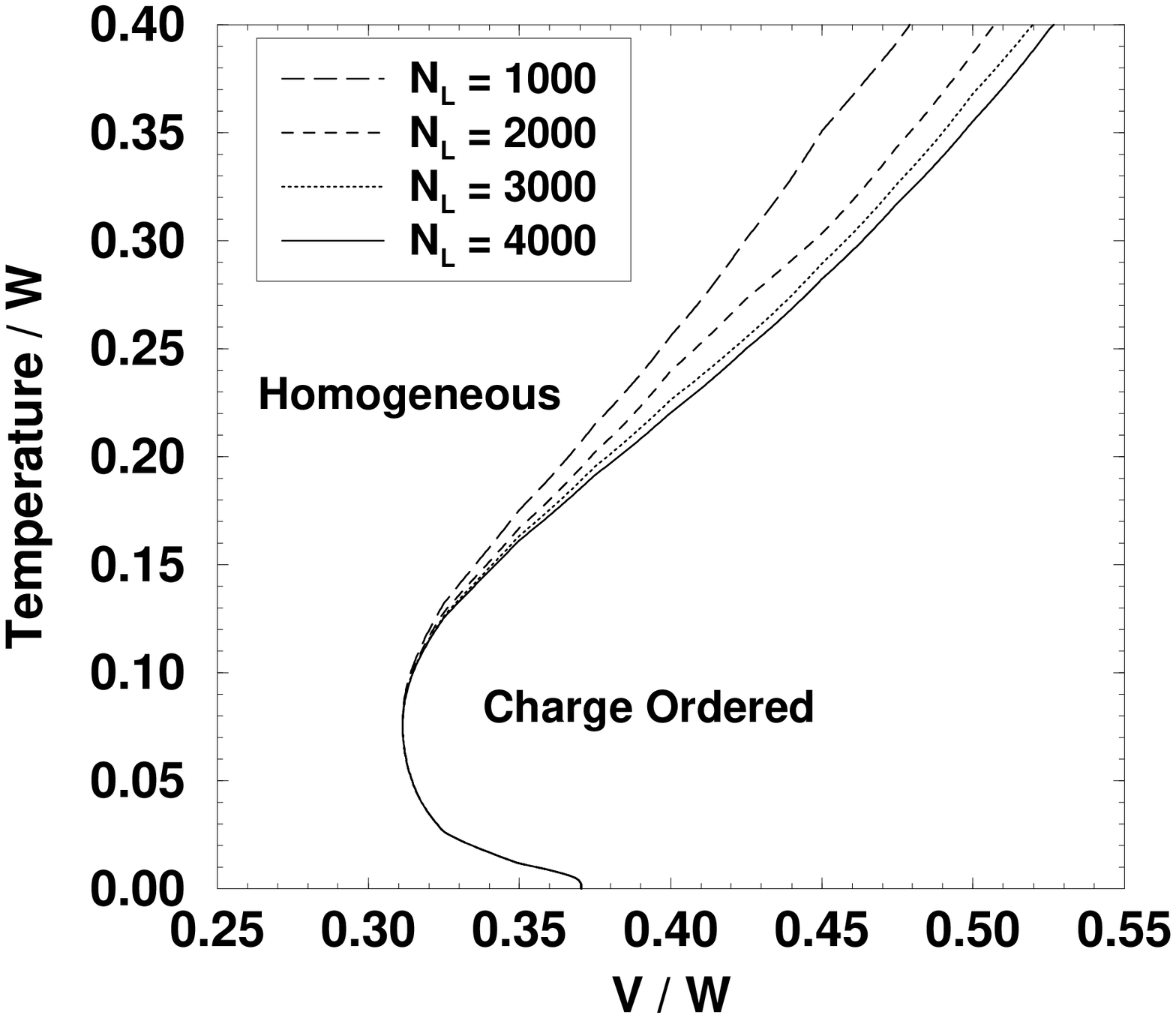}}
\vspace{.1in}
\caption{Phase diagram for the model with $\chi_c=0.1$.
There is clear reentrant behavior around
$V/W \approx 0.35$.
The four curves correspond to different numbers of Lanczos steps.
Runs with more Lanczos steps result in more eigenvalues, and thus
are accurate to higher temperatures.
The N$_{\rm L}=4000$ run is clearly accurate well into the 
high-temperature homogeneous phase in the reentrant region.
}
\label{fig:phase}
\end{figure}

The phase diagram calculated from the point
where $\chi(V,T) = \chi_c = 0.1$ is shown in Fig.\ \ref{fig:phase}.
Reentrant behavior is seen for $0.31 \lesssim V/W \lesssim 0.37$.
For $V$ in this range, the ground state is homogeneous,
but a CO phase exists at intermediate temperatures.

The 4$\times$4 system has no entropy in either the homogeneous
or the CO phases, so the phase boundary in Fig.\ \ref{fig:phase}
has infinite slope in the $T \rightarrow 0$ limit.
However, the CO phase has more low-lying excitations than
the homogeneous phase, so in the reentrant region the CO
phase becomes favored at intermediate temperatures.

The greater number of low-lying excitations in the CO phase 
may be at least partially due to the topology of the 4$\times$4 periodic
cluster used.
The homogeneous phase will have the usual Fermi-liquid excitations.
In the CO phase, the electrons order in a checker-board pattern.
The spins interact via a fourth order superexchange process \cite{goodenough63,manousakis91a}.
A given spin can interact both with its diagonal neighbor
and with its neighbor two steps away either horizontally or vertically.
On the infinite lattice, the interaction with the diagonal neighbor
is 4 times as strong as the interaction with the horizontal/vertical
neighbor because there are 4 times as many diagonal superexchange paths.
Thus the spin system is equivalent to the extended
antiferromagnetic Heisenberg 
model on a square lattice with $J_1 = 4 J_2$, where 
$J_1$ is the nearest-neighbor interaction and
$J_2$ is the next-nearest-neighbor interaction.
Extensive numerical calculations have shown that this Heisenberg
model forms a N\'eel state at $T=0$ \cite{schulz96a,singh99a}.

Because two steps to the left is the same as two steps to the right (and
the same for up and down),
the 4$\times$4 periodic cluster is equivalent to a hypercubic $2^4$ cluster
\cite{dagotto90a}.
Thus the diagonal superexchange coupling is the same as the 
horizontal/vertical coupling.
So the CO N\'eel state on this particular cluster is strongly frustrated
and may not form.
This could contribute to the greater density of low lying states
in the CO phase than in the homogeneous phase.

Smaller periodic clusters have even more severe finite-size problems.
We decided to work with the 4$\times$4 cluster because it is the largest
system with a manageable Hilbert space.
Even with all symmetries applied, the largest
Hamiltonian matrices were larger than $2.0 \times 10^5$
(the matrices are very sparse, with only about 30 non-zero elements
in each row or column).
The Hilbert space diverges exponentially with the number of sites in
the cluster.
The next largest cluster with an even number of electrons
has 20 sites and has irreducible sectors with Hilbert spaces larger 
than $2.4 \times 10^8$, more than 1000 times larger than those of 
the 16-site system.

In summary, we have examined the reentrant charge order transition
in a simple model for the manganites, the extended Hubbard model
in two dimensions.
We computed the low-lying eigenvalues of the model on a 16-site periodic
cluster, and used the susceptibility with respect to the nearest-neighbor
repulsion to compute the low-temperature phase boundary between
the homogeneous and charge-ordered phases.
Like previous results on this model in infinite dimensions, we find
a parameter region where the model shows reentrant behavior.
The reentrant region may be partially an artifact of the small cluster used,
which frustrates the formation of N\'eel spin order in the charge ordered state.

This work was supported by the Office of Naval Research.
Computations were performed on the SGI Origin 2000 at the ASC
DoD Major Shared Resource Center.


\end{document}